\documentclass[useAMS,usenatbib,usegraphicx]{mn2e}

\title{On the physical origin of dark matter density profiles}
\author[Y.~Ascasibar et al.]
{Y.~Ascasibar,$^{1, 2, 3}$ G.~Yepes,$^2$ S.~Gottl\"ober$^3$ and V.~M\"uller$^3$\\
 $^1$Theoretical Physics, 1Keble Road, Oxford OX1 3NP\\
 $^2$Grupo de Astrof\'\i sica, Universidad Aut\'onoma de Madrid, Madrid E-28049, Spain\\
 $^3$Astrophysikalisches Institut Potsdam, An der Sternwarte 16, Potsdam D-14482, Germany
}
\def\LCDM{$\Lambda$CDM}
\def\lcdm{\LCDM\ }
\def\Om{\Omega_{\rm m}}
\def\OL{\Omega_\Lambda}

\newcommand{\ii}{_{\rm i}}
\newcommand{\ff}{_{\rm f}}
\newcommand{\mm}{_{\rm m}}
\newcommand{\sss}{_{\rm s}}

\def\Am{\alpha_{\rm M}}
\def\Madd{M_{\rm add}}
\def\Rv{r_{\rm vir}}
\def\Vv{v_{\rm vir}}

\newcommand{\be}{\begin{equation}}
\newcommand{\ee}{\end{equation}}
\newcommand{\bea}{\begin{eqnarray}}
\newcommand{\eea}{\end{eqnarray}}

\newcommand{\dd}{{\rm d}}

\begin{document}

\maketitle

\begin{abstract}
The radial mass distribution of dark matter haloes is investigated
within the framework of the spherical infall model.
We present a new formulation of spherical collapse including
non-radial motions, and compare the analytical profiles with a set of
high-resolution N-body simulations ranging from galactic to cluster
scales.
We argue that the dark matter density profile is entirely determined
by the initial conditions, which are described by only two parameters:
the height of the primordial peak and the smoothing scale. These are
physically meaningful quantities in our model, related to the mass and
formation time of the halo.
Angular momentum is dominated by velocity dispersion, and it is
responsible for the shape of the density profile near the centre.
The phase-space density of our simulated haloes is well described by a
power-law profile,
$\rho/\sigma^3 = 10^{1.46\pm0.04} (\rho_c/\Vv^3) (r/\Rv)^{-1.90\pm0.05}$.
Setting the eccentricity of particle orbits according to the numerical
results, our model is able to reproduce the mass distribution of
individual haloes.
\end{abstract}

\begin{keywords}
galaxies: haloes -- cosmology: theory -- dark matter
\end{keywords}

%--------------------------------------------------------------------------
  \section{Introduction}
  \label{secIntro}
%--------------------------------------------------------------------------

The hierarchical clustering paradigm states that the growth of
cold dark matter (CDM) haloes proceeds by accretion of smaller units
from the surrounding environment, either by continuous infall or by
discrete merging events.

Cosmological N-body simulations are a valuable tool to study the
mass distribution of dark matter haloes and its evolution in the
non-linear regime.
In the early numerical work of \citet{QSZ86} and
\citet{Frenk88}, haloes showed an isothermal
density profile ($\rho\propto r^{-2}$).
\citet{DubinskiCarlberg91} and
\citet{Crone94} had enough resolution in their simulations
to detect the first evidence of departure from a pure power-law.
Later on,
\citet[hereafter NFW]{NFW96,NFW97} found that the density profile
could be fitted by a simple analytical function
\be
\rho(r)=\frac{\rho\sss}{(r/r\sss)(1+r/r\sss)^2}
\label{ecNFW}
\ee
in terms of a characteristic density $\rho\sss$ and a characteristic
radius $r\sss$. This profile is steeper than isothermal at large
radii and shallower near the centre.
The logarithmic slope of the density profile,
$\alpha(r)\equiv\dd\log(\rho)/\dd\log(r)$, tends to $\alpha=-3$ for
$r\to\infty$ and $\alpha=-1$ for $r\to0$.
It corresponds to the isothermal case, $\alpha=-2$, at the
characteristic radius only.
\cite{NFW97} further showed that the two free parameters in equation
(\ref{ecNFW}) are not independent.
Should this be true, the final mass distribution of objects of
different scales could be described in terms of a one-parameter family
of analytical profiles.

Similar results have been found in independent simulations with
much higher mass and force resolution than the original NFW
paper. However, there is still some controversy about the innermost
value of $\alpha$ and its dependence on resolution.
\citet{Moore98,Moore99}, \citet{Ghigna98,Ghigna00} and
\citet{FukushigeMakino97,FukushigeMakino01} find steeper density
profiles near the centre ($\alpha\sim-1.5$), whereas other authors
\citep{JingSuto00,Klypin01} claim that the actual value of $\alpha$
may depend on halo mass, merger history, and substructure.
\citet{Power03} pointed out that the
logarithmic slope becomes increasingly shallow inwards, with little
sign of approaching an asymptotic value at the resolved radii. In that
case, the precise value of $\alpha$ at a given cut-off scale
would not be particularly meaningful.
This result has been later confirmed by \citet{Fukushige_03} and
\citet{Hayashi_03}, and it is predicted by several analytical
models \citep[e.g.][]{TN01,Hoeft_03}.

Observed rotation curves of dwarf spiral and LSB galaxies
\citep[e.g.][]{FloresPrimack94,Moore94,Burkert95,KKBP98,BorrielloSalucci01,Blok01,BlokBosma02,Marchesini02}
seem to indicate that the shape of the density profile at small scales
is significantly shallower than what is found in numerical
simulations. This discrepancy has been often signalled as a genuine
crisis of the CDM scenario, and several alternatives have been
suggested, such as warm \citep{Colin00,SommerLarsenDolgov01},
repulsive \citep{Goodman00},
fluid \citep{Peebles00},
fuzzy \citep{Hu00},
decaying \citep{Cen01},
annihilating \citep{Kaplinghat00},
or self-interacting \citep{SpergelSteinhardt00,Yoshida00,Dave01}
dark matter.

Unfortunately, it has proved remarkably hard to establish the inner
slope of the dark matter distribution observationally \citep[see
e.g.][]{Swaters03}. Some authors
\citep{BoschSwaters01,JimenezVerdeOh03,Swaters03}
claim that a cuspy density profile with $\alpha\leq-1$ is consistent
with current observations, although a shallower slope is
able to explain them as well. Yet, a value as steep as $\alpha=-1.5$
can be confidently ruled out in most cases.
According to \citet{Hayashi_03}, only about 30 per cent of the
rotational curves are actually inconsistent with the simulation data.

On cluster scales, X-ray analyses have led to wide ranging results,
from $\alpha=-0.6$ \citep{EttoriFAJ02} to $\alpha=-1.2$ \citep{Lewis03}
or even $\alpha=-1.9$ \citep{Arabadjis02}.
Measurements based on gravitational lensing yield conflicting
estimates as well, either in rough agreement with the results of
numerical simulations \citep[e.g.][]{Dahle03,Gavazzi03}, or finding
much shallower slopes, $\alpha=-0.5$ \citep[e.g.][]{Sand02,Sand_03}.

A conclusive theoretical prediction of the central mass distribution
of CDM haloes is therefore an important check for any model of structure
formation. The controversy regarding the 'universal' density profile
and its logarithmic slope at the centre has
stimulated a great deal of analytical work.
On one hand, we would like to find out not only the actual shape of
the profile, but also the physical mechanisms behind the
'universality' observed in numerical N-body simulations.
On the other hand, it would be interesting to find and explain
correlations with  halo size, environment, power
spectrum or even the nature of dark matter particles.

A number of plausible arguments about the radial structure of CDM
haloes have been advanced during the last 30 years.
The basic problem of the collisionless collapse of a spherical
perturbation in an expanding background was first addressed in the two
seminal papers by \citet{GG72} and \citet{Gunn77}, where the
cosmological expansion and the role of adiabatic invariance were first
introduced in the context of the formation of individual objects.
The next step was accomplished by \citet{FG84} and
\citet{Bertschinger85}, who found analytical predictions for the
density of collapsed objects seeded by scale-free primordial
perturbations in a flat universe. \citet{HS85} generalised these
solutions to realistic initial conditions in flat as well as open
Friedmann models. Modifications of  the self-similar collapse model to
include more realistic dynamics of the growth process have been
proposed
\citep[e.g.][]{AFH98,HenriksenWidrow99,Lokas00,Kull99,SCO00}. Several
authors \citep[e.g.][]{syerWhite98, Salvador98, NusserSheth99,
  Manrique03} argue that the central density profile is linked to the
merging history of dark matter substructure, and baryons have been
invoked both to shallow \citep[e.g.][]{ElZant01,ElZant_03}
and to steepen \citep{Blumenthal86} the dark matter profile.

In this paper, we present an analytical model for the assembly of CDM haloes
based on the spherical collapse paradigm.
Following the spirit of \citet{HS85},
we will assume that objects do not form around \emph{local} maxima
of the primordial density field, but of the \emph{smoothed} density field.
We argue that all information about the initial conditions
below the smoothing scale is lost during the merging process.
We will show that setting the smoothing scale for a halo of a given mass
is equivalent to specifying its formation time.
Initial conditions are computed following the
Gaussian random peaks statistics described by \citet[hereafter
BBKS]{BBKS86}, and angular momentum is included in a phenomenological way.
Our theoretical predictions are then compared with the results of
high-resolution numerical simulations, showing that this
model is able to accurately reproduce the mass distribution of
individual objects.

The paper is structured as follows.
Details of our implementation of spherical collapse are given in
Section~\ref{secSC}.
Numerical experiments are described in Section~\ref{secSims}, where
both the mass and velocity distributions are investigated.
We discuss our results in Section~\ref{secComp}, and
Section~\ref{secConclus} summarises our main conclusions.

%--------------------------------------------------------------------------
  \section{Spherical collapse}
  \label{secSC}
%--------------------------------------------------------------------------

The assembly of dark matter haloes is a highly non-linear process, and strong simplifying assumptions must be made in order to tackle the problem analytically. Traditionally, there are two complementary paradigms: the spherical infall model \citep{GG72} and the Press-Schecter formalism \citep{PS74}, in which mergers play a dominant role \citep[see e.g.][]{NusserSheth99,Manrique03}.

%___________________________________________________

\subsection{The model}
\label{secSCmodel}

The most simple way of addressing the problem of structure formation
is to assume spherical symmetry.
As shown by \citet{Tolman34} and \citet{Bondi47}, a spherically
symmetric solution of the Einstein equations can be easily interpreted 
in terms of Newtonian dynamics. The equation of
motion for a Lagrangian shell enclosing a mass $M$ can be derived from
the conservation of energy
\be
\epsilon(r)\equiv\frac{E(r)}{m}\equiv
\frac{\dot r^2}{2}-G\frac{M(r)+\frac{4\pi}{3}\rho_\Lambda r^3}{r}
=\epsilon\ii(r)
\label{ecConservE}
\ee
where $r(t)$ is the Lagrangian coordinate, $M(r)=M\ii(r\ii)$ is the mass
contained within the shell, and
$\rho_\Lambda\equiv\frac{\Lambda}{8\pi G}$ is the vacuum energy
density associated to a cosmological constant.
Throughout this paper, the subscript 'i' will be used to
denote initial conditions.

If we assume the universe to be homogeneous, the mass is given by
$M\ii=\frac{4\pi}{3}\Om^i\rho_c^i r\ii^3$, where
$\rho_c^i\equiv\frac{3H\ii^2}{8\pi G}$. Substituting in
(\ref{ecConservE}) and making a coordinate transformation
\be
r(t)\equiv r\ii\alpha(t)
\ee
we obtain the Friedmann equation without radiation:
\be
\frac{\dot \alpha^2}{H\ii^2}-\Om^i \alpha^{-1}-\OL^i \alpha^2
= 1-\Om^i -\OL^i.
\ee

In a homogeneous universe, $\alpha(t)$ is independent of $r\ii$ and
plays the role of a uniform cosmic expansion factor.
Our model assumes that structures grow from spherically symmetric
perturbations, defined as
\be
\Delta\ii(r\ii)\equiv\frac{M(r\ii)}{\frac{4\pi}{3}\Om^i\rho_c^i r\ii^3}-1
\ee
with $0<\Delta\ii\ll1$.
These perturbations are introduced at an early epoch $a\ii$
by slightly displacing the spherical shells of matter.
Keeping only terms to first order in $\Delta\ii(r\ii)$,
the new positions and velocities are given by
\be
r\ii'\simeq r\ii\left(1-\frac{1}{3}\Delta\ii(r\ii)\right)~~;~~
v\ii'\simeq H\ii r\ii\left(1-\frac{2}{3}\Delta\ii(r\ii)\right).
\ee

The mass enclosed by the perturbed shell is still $M\ii$, and the initial
specific energy (using $\Om^i\simeq 1$ and $\OL^i\simeq 0$)
is $\epsilon\ii\simeq-\frac{5}{6}\Delta\ii(H\ii r\ii)^2$.
Energy conservation (\ref{ecConservE}) leads to the equation of motion
\be
-\frac{5}{3}\Delta\ii\simeq
\frac{\dot \alpha^2}{H\ii^2}-\Om^i \alpha^{-1}-\OL^i \alpha^2.
\label{ecSC}
\ee

According to this equation, the evolution of a single spherical shell
would be similar to that of a Friedmann universe. For high enough
values of $\Delta\ii$, the shell reaches a maximum radius $r\mm$ at a
{\em turn-around} time $t\mm$ and then re-collapses. In an Einstein-de
Sitter universe ($\Om=1$, $\OL=0$), equation (\ref{ecSC}) can be solved
analytically, yielding
\be
r\mm=\frac{3r\ii}{5\Delta\ii} ~~;~~ t\mm=\frac{\pi}{2H\ii(5\Delta\ii/3)^{3/2}}.
\label{ecRmTm}
\ee

This is also a valid approximation for shells reaching their
turn-around radius $r\mm$ before the cosmological constant term starts
to dominate the expansion.
Since the shells need at least another $t\mm$ to virialise, expression
(\ref{ecRmTm}) can be applied in a \lcdm universe to estimate the
maximum expansion radius and time for the innermost part of a
virialised halo.
For the outer shells, the equation of motion (\ref{ecSC})
must be integrated numerically in order to find the trajectories up to 
the maximum radius.

In the absence of shell-crossing, shells would reach the origin at
$T=2t\mm$. Since they are assumed to be composed of collisionless CDM
particles, they would simply pass through the centre, describing an
oscillatory motion with amplitude $r\mm$ and period $T$.
However, equation (\ref{ecSC}) holds as long as the enclosed mass
$M\ii$ remains constant. As a shell re-collapses, its particles will
cross the orbits of inner shells, and energy will not be a constant of
motion any more.

After turn-around, our model assumes that the particles belonging to a
shell oscillate (or, more generally, {\em orbit}) within the
gravitational potential of the dark matter halo.
Since CDM particles are expected to spend most of the time in the
outermost regions of their orbits (particularly when angular momentum
is taken into account), we approximate the mass distribution of the
halo by a simple power law
\be
M(r)=M\ii\left(\frac{r}{r\mm}\right)^{\Am(r\mm)}
\label{ecSCansatz}
\ee
where $\Am(r)\equiv \dd \log M(r)/\dd \log r$ is the {\em local} value
of the logarithmic slope of the mass profile, evaluated at the maximum 
radius of the orbit.

At first sight, this might seem similar to the classical approach
based on self-similarity \citep{Bertschinger85},
but in that case the final mass distribution is indeed assumed to be a
power law, whereas in our model this ansatz is only an approximation
to compute the local potential. The final mass profile is obtained
self-consistently, adding the contributions from all shells, each of
them with an individual value of $\Am(r\mm)$.

The probability of finding a particle inside radius $r$ is
proportional to the fraction of time it spends within $r$:
\be
P(r,r\mm) = \frac{1}{t\mm} \int_0^r\frac{\dd x}{v_r(x)} =
\frac{1}{t\mm} \int_0^r\frac{\dd x}{\sqrt{\Phi(r\mm)-\Phi(x)}}.
\label{ecPr}
\ee

We evaluate numerically the value of $\Am(r\mm)$ in order to compute
the potential. Taking different prescriptions, such as an isothermal
profile ($\Am=1$ for every shell) or a Keplerian potential ($\Am=0$)
does not lead to qualitative variations in the probability $P(r,r\mm)$
and the resulting mass distribution.

If phase mixing is considered to be efficient, particles initially on the same shell will be spread out from $r=0$ to $r=r\mm$ a short time after $t\mm$. For the sake of simplicity, we will consider that phase mixing is instantaneous, so immediately after turn-around the shell is transformed into a density distribution whose cumulative mass is proportional to $P(r,r\mm)$.

This means that recently accreted particles
contribute to the mass within $x\mm<r\mm$ (i.e. shell-crossing). For
shells whose maximum radius was $x\mm$, the enclosed mass changes from
$M\ii(x\mm)$ to 
\be
M(x\mm)=M\ii(x\mm)+\Madd(x\mm)
\ee
where $\Madd(x\mm)$ accounts for particles belonging to outer
shells. To compute $\Madd(x\mm)$ \citep[see][]{ZH93}, we must
integrate the contribution of every shell whose maximum radius is
larger than $x\mm$, up to the current turn-around radius $R_{\rm ta}$:
\be
\Madd(x\mm) =
\int_{x\mm}^{R_{\rm ta}} {\frac{\dd M\ii(r)}{\dd r} ~ P(x\mm,r) ~ \dd r}.
\ee

To compute the evolution of the shell after shell-crossing, we apply adiabatic invariance \citep{Gunn77}.
If the potential evolves on a timescale much longer than the orbital period of the inner particles, their dynamics admits an adiabatic invariant
\be
J_r=\frac{1}{2\pi}\oint v_r(r)~\dd r
\ee
also known as the radial action. For a power-law potential,
the radial action $J_r$ is proportional to $\sqrt{x\mm M(x\mm)}$.
When we increase the mass by an amount $\Madd$, the maximum radius of
the inner shell must decrease in order to keep $J_r$ constant.
The final radius, $x_0$, is given by the implicit equation
$x_0=F(x_0)\ x\mm$, where
\be
F(x_0)=\frac{M\ii}{M\ii+\Madd(x_0)}=\frac{M\ii}{M\ii+\Madd[F(x_0)x\mm]}
\label{ecFr}
\ee
and whose solution must be obtained numerically for each shell.

To summarise, the numerical procedure to compute the final
radius $r_0(t_0)$ of a Lagrangian shell of matter involves
the following steps:
\begin{enumerate}
  \item Set $M\ii$ (or, equivalently, $r\ii$). Start by the outer shell.
  \item Integrate the equation of motion (\ref{ecSC}) up to $t\mm$.
  \item If $t\mm>t_0$, the shell is still expanding: $r_0=r(t_0)$.
  \item If $t\mm<t_0$, solve (\ref{ecFr}) to compute $r_0=F(r_0)r\mm$, and add the contribution of this shell to $\Madd(r)$.
  \item Repeat for the next shell towards the centre.
\end{enumerate}

%___________________________________________________

\subsection{Initial conditions}

The model described above allows us to compute the mass
distribution arising from a primordial fluctuation $\Delta\ii(r\ii)$,
but does not say anything about the shape of this function or its
physical origin. Never the less, it is important to note that the
final density profile is entirely determined by this initial
condition. Thus, in the spherical collapse paradigm, the case for a
universal density profile can be reformulated in terms of universality
in the primordial fluctuations that set $\Delta\ii(r\ii)$.

\citet{HS85} suggested that, according to the hierarchical scenario of
structure formation, haloes should collapse around maxima of the
{\em smoothed} density field. The statistics of peaks in a Gaussian
random field has been extensively studied in the classical paper by BBKS.
A well-known result is the
expression for the radial density profile of a fluctuation centered on
a primordial peak of arbitrary height:
\be
\frac{\left<\delta\ff(r)\right>}{\sigma_0} = \nu\psi(r)-
\frac{\theta(\gamma,\gamma\nu)}{\gamma(1-\gamma^2)}
\left[ \gamma^2\psi(r) + \frac{R_*^2}{3}\nabla^2\psi(r) \right]
\label{ecBBKS}
\ee
where
$\psi(r)\equiv\xi(r)/\sigma_0$ is the normalised two-point correlation
function,
$\sigma_0\equiv\xi(0)^{1/2}$ is the rms density fluctuation,
and $\nu\sigma_0$ is the height of the peak.
The quantities $\gamma\equiv\sigma_1^2/(\sigma_2\sigma_0)$ and
$R_*\equiv\sqrt{3}\sigma_1/\sigma_2$ are related to the moments of the
power spectrum,
\be
\sigma_j^2 \equiv \frac{1}{2\pi^2}\int_0^\infty{P(k)~k^{2(j+1)}~\dd k},
\ee
and the function $\theta(\gamma,\gamma\nu)$ parametrizes the second
derivative of the density fluctuation near the peak. BBKS suggest the
approximate fitting formula
\be
\theta(\gamma,\gamma\nu) \simeq
\frac{3(1-\gamma^2)+(1.216-0.9\gamma^4)\exp[-\frac{\gamma}{2}(\frac{\gamma\nu}{2})^2]}
     { \left[ 3(1-\gamma^2)+0.45+\frac{\gamma\nu}{2} \right]^{1/2} + \frac{\gamma\nu}{2}}
\ee
valid to 1 per cent accuracy in the range $0.4<\gamma<0.7$ and
$1<\gamma\nu<3$, which is the scale relevant for both galaxies and
galaxy clusters.

Expression (\ref{ecBBKS}) is often quoted in the literature
\citep[e.g.][]{Hoffman88,LokasHoffman00,Popolo00,Hiotelis02} as the
initial condition $\Delta\ii(r\ii)$.
However, we argue that $\left<\delta\ff(r)\right>$ denotes the {\em
  smoothed} density profile around a local maximum of the smoothed
density field,
\be
\delta\ff(\vec r)=\int{ W\ff(\vec r-\vec x)\delta(\vec x) \dd^3\vec x}
\label{ecSmooth}
\ee
where the function $W\ff(\vec r-\vec x)$ is a smoothing kernel that depends on a certain filtering scale $R\ff$.

The smoothed profile $\left<\delta\ff(r)\right>$ given by
(\ref{ecBBKS}) is in general not equal to the mean value of the actual
overdensity, $\delta(r)$, that must be integrated to
compute $\Delta\ii(r\ii)$:
\be
\Delta\ii(r\ii) = 4\pi \int_0^{r\ii} \delta(\vec x) x^2 \dd x.
\label{ecDelta0}
\ee

Comparing this expression with (\ref{ecSmooth}), we see that
$\Delta\ii(r\ii)$ is equivalent to $\delta\ff(0)$ as long as $W\ff$ is
taken to be a spherical top hat of radius $r\ii$.

To sum up, we are interested in the {\em physical} density profile
$\Delta\ii(r\ii)$ around a local maximum of the {\em smoothed} density field.
To locate the maximum, we use a Gaussian smoothing kernel
\be
W\ff(\vec r-\vec x)=
{(2\pi R\ff^2)}^{-3/2}
\exp{\left(-\frac{|\vec r-\vec x|^2}{2R\ff^2}\right)}
\ee
in order to avoid the oscillations in Fourier space that arise from
a top hat filter.
We set the scale of the fluctuation, $R\ff$, and impose the condition
that $r=0$ is a maximum of $\delta\ff(\vec r)$.

Then, we compute $\Delta\ii(r\ii)=\delta_{r\ii}(0)$
by applying a top hat smoothing of radius $r\ii$.
BBKS show that the
probability distribution of $\delta_{r\ii}(0)$ is a Gaussian with mean
\be
\left< \delta_{r\ii}(0) \right> =
 \nu\ff \frac{\sigma_{0h}^2} {\sigma_{0f}} -
 \frac{\gamma\ff\theta(\gamma\ff,\nu\ff)} {1-\gamma\ff^2}
 \frac{\sigma_{0h}^2} {\sigma_{0f}}
\left( 1-
   \frac{\sigma_{1h}^2~\sigma_{0f}^2} {\sigma_{0h}^2~\sigma_{1f}^2}
\right).
\ee
The moments
\be
\sigma_{jx}^2\equiv \frac{1}{2\pi^2}\int_0^\infty{P_x(k)~k^{2(j+1)}~\dd k}
\ee
are computed from
\be
P\ff(k)\equiv P(k)~\exp\left[-(kR\ff)^2\right]
\ee
and
\be
P_h(k)\equiv P(k)~
\exp\left[-\frac{(kR\ff)^2}{2}~\right]
\frac{3[\sin(kr\ii)-kr\ii\cos(kr\ii)]}{(kr\ii)^3}
\ee
where $P(k)$ is the ${\Lambda{\rm CDM}}$ power spectrum that we
used to generate the initial conditions for our simulations, evaluated
at time $a\ii$.

%___________________________________________________

\subsection{Angular momentum}
\label{secSCj}

Two decades after the seminal paper by \citet{GG72},
\citet{WhiteZaritsky92} pointed out \citep[see also the pioneering
work of][]{GurevichZybin88} that angular momentum would prevent the
orbits of CDM particles from reaching the origin.

For pure radial orbits, the mass in the centre is dominated by $\Madd$
(i.e. particles from the outer shells) when the profile at turn-around
is shallower than isothermal \citep{FG84}. The final density
distribution is $\rho(r)\propto r^{-2}$, independent on the initial
logarithmic slope. More recently, \citet{Lokas00} and
\citet{LokasHoffman00} found a similar result considering
non-self-similar primordial fluctuations based on the peak
formalism.

Angular momentum arises in linear theory \citep{White84} from the
misalignment between the asymmetry of the collapsing region (i.e. the
inertia tensor) and the tidal forces it experiences during the linear
regime. In addition, violent relaxation \citep{LB67} transforms the
radial infall energy into random velocity dispersion, which has a
tangential component.

The total amount of angular momentum acquired by the system is, however, an open question. The usual approach \citep{AFH98,Nusser01,Hiotelis02} consists in assigning a specific angular momentum at turn-around
\be
j\propto\sqrt{GMr\mm}.
\ee

With this prescription, the orbit eccentricity $e$ is the same for all
particles in the halo \citep{Nusser01}.
The pericentric radius is given by
\be 
r_{\rm min}=\frac{1-e}{1+e}~r_{\rm max}
\label{ecRmin}
\ee
where $r_{\rm max}$ is computed from adiabatic invariance
(\ref{ecFr}), following the procedure explained in
Section~\ref{secSCmodel}. Angular momentum is taken into account by
adding the usual term $j^2/(2r^2)$ to the gravitational potential
$\Phi(r)$ in equation (\ref{ecPr}). Although this changes the actual
value of the radial action, $J_r$ is still proportional to $\sqrt{r\mm 
  M(r\mm)}$ and (\ref{ecFr}) can be used to compute the collapse factor $F(r)$. The assumption of spherical symmetry leads to angular momentum conservation, which implies constant $e$ during the contraction.

%__________________________________
\begin{figure}
  \centering \includegraphics[width=7cm]{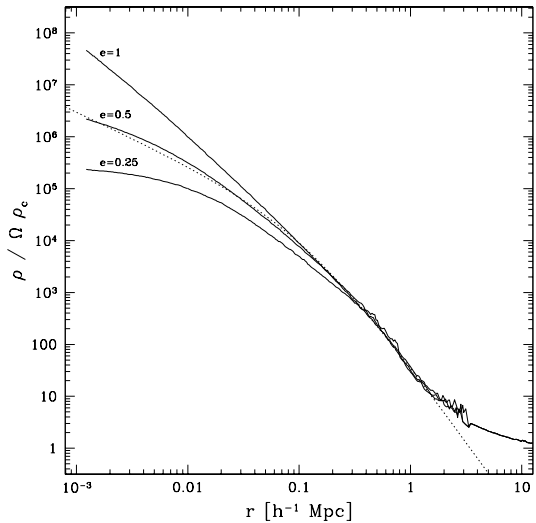}
  \caption{Radial density profile arising from a $3\sigma$ fluctuation
    on $1\ h^{-1}$ Mpc scales (solid lines), for different values of
    the orbit eccentricity $e$.
    A NFW profile (dotted line) is plotted for comparison.}
  \label{figSCj}
\end{figure}
%__________________________________

The effect of angular momentum is illustrated in Figure~\ref{figSCj}. We
plot the mass distribution at $z=0$ corresponding to a $3\sigma$ peak
in the primordial density field, smoothed on $1\ h^{-1}$ Mpc
scales. As noted by \citet{WhiteZaritsky92} and \citet{Hiotelis02},
angular momentum plays a key role during secondary infall, decreasing
the amount of mass contributed by recently accreted shells
in the innermost regions of the halo.

In agreement with \citet{Hiotelis02}, we find that
the predicted density profile becomes shallower in the centre
as the amount of angular momentum is increased (lower $e$).
Pure radial orbits give rise to a steep profile, somewhat similar to
the form proposed by \citet{Moore99}, although the exact
shape is slightly different.
When the eccentricity is low, the halo develops a constant density
core inconsistent with the results of numerical simulations.
A value $e=0.5$ yields a mass distribution virtually
indistinguishable from the NFW formula (dotted line in Figure~\ref{figSCj})
down to kpc scales.
However, the logarithmic slope predicted by the model keeps increasing
towards the centre, while NFW tends to an asymptotic value, $\alpha=-1$.

%--------------------------------------------------------------------------
  \section{Numerical Experiments}
  \label{secSims}
%--------------------------------------------------------------------------

In order to test the analytical model,
we carried out a series of high-resolution N-body simulations of
cluster formation with the adaptive mesh code ART \citep{ART97}.
Slightly lower spatial resolution experiments have been run from the same
initial conditions with the Tree-SPH gasdynamical code {\sc Gadget}
\citep{Gadget01,GadgetEntro02}. The radial structure of both dark and
baryonic components has been addressed in \citet{Ascasibar03}.
For a detailed description of the numerical experiments, the reader is
referred to \citet{tesis}.

A sample of cluster-size haloes has been selected from an initial
low-resolution ($128^3$ particles) pure N-body simulation of a 80
$h^{-1}$ Mpc box in a \lcdm universe ($\Omega_{\rm m}=0.3$;
$\Omega_\Lambda=0.7$; $ h=0.7$; $\sigma_8=0.9$). Higher resolution has
been achieved
by means of the multiple-mass technique \citep[see][for
details]{Klypin01}, using 3 levels of mass refinement.
This is equivalent to an effective resolution of $512^3$ CDM particles
($3.1\times 10^8\ h^{-1}$ M$_\odot$) in the highest refinement level.
The minimum cell size allowed in the ART runs was $1.2\ h^{-1}$ kpc.
All simulations were started at $z=50$.

This procedure has been applied to 15 objects, ranging from
$3\times10^{13}$ to $2\times 10^{14}\ h^{-1}$ M$_\odot$. In order to
explore a broader mass range, a smaller box (25 $h^{-1}$ Mpc) has been 
simulated, and six galaxy-size haloes have been added to the cluster
sample. Mass resolution is $1.2\times 10^6\ h^{-1}$ M$_\odot$ for
these objects, with a minimum cell size of 0.2 $h^{-1}$ kpc.

All CDM haloes have been classified according to their dynamical state 
according to a substructure-based criterium.
We use the {\sc Bound Density Maxima} galaxy finding algorithm \citep[see
e.g.][]{Colin99,Klypin99} and look for the most massive subhalo
within the virial radius. We label as {\em major merger} any object
in which the subhalo is heavier than 50 per cent the mass of the
main object; if the mass is between 10 and 50 per cent, it is
classified as a {\em minor merger}; otherwise, we assume the object to
be a {\em relaxed} system in virial equilibrium.

%___________________________________________________

\subsection{Mass distribution}
 \label{secDM}

\begin{figure}
  \centering \includegraphics[width=8cm]{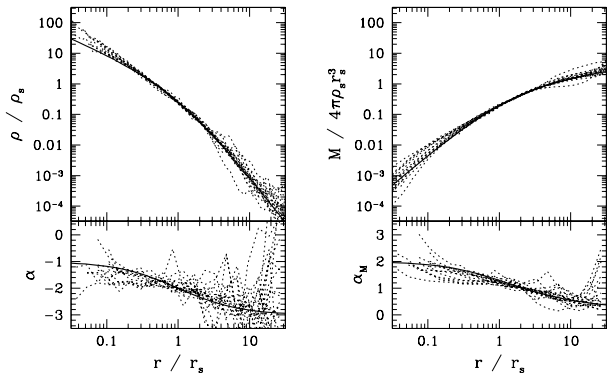}
  \caption{Radial density (left) and mass (right) profiles of our dark
    matter haloes, rescaled by their best-fitting NFW parameters and
    plotted as dotted lines.
    The corresponding logarithmic slopes are shown on the bottom
    panels.
    NFW model is indicated by a solid line.
    }
  \label{figDMprof}
\end{figure}

The spherically-averaged mass distribution of our numerical CDM haloes
is shown in Figure~\ref{figDMprof}. All profiles have been rescaled
by their best-fitting characteristic density and radius.
As can be seen in the Figure, the NFW formula provides a reasonable
approximation to the mass distribution, although some deviations occur 
at the innermost part, as well as beyond the virial radius.

%__________________________________
%__________________________________

\begin{figure}
  \centering \includegraphics[width=7cm]{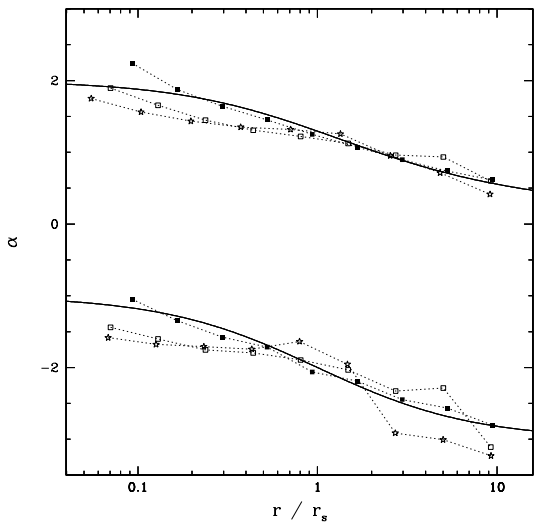}
  \caption{Logarithmic slopes of the mass (top) and density (bottom)
    profiles.
    Black squares show the average over relaxed haloes.
    Minor and major mergers are represented by empty squares and
    stars, respectively.
    Solid lines indicate NFW model.
}
\label{figSlope}
\end{figure}

In Figure~\ref{figSlope}, the logarithmic slopes of both mass and
density profiles
are plotted according to our dynamical classification.
We do not find any evidence of an asymptotic behavior
up to the resolution limit, in agreement with recent numerical studies 
\citep{Power03,Fukushige_03,Hayashi_03}.
An extrapolation of our results suggests that the central
slopes in relaxed haloes could be indeed less steep than the NFW
fit, as predicted by analytical models based on the velocity
dispersion profile \citep{TN01,Hoeft_03}. More resolution is clearly
needed in order to establish a firm conclusion.

%__________________________________
%__________________________________

As pointed out by several authors
\citep[e.g.][]{JingSuto00,Klypin01,Ascasibar03}, the mass distribution 
near the centre might depend on the dynamical state of the halo.
At the resolution of the present simulations, relaxed haloes are well
described by NFW model, but merging systems display steeper profiles.
A conjecture that deserves further investigation is whether violent
relaxation and subsequent equipartition of energy may drive the inner
part of the system into a nearly isothermal state.
That would provide a natural mechanism for the final profile to be
independent on the details of the merging history.

%__________________________________
\begin{figure}
  \centering \includegraphics[width=7cm]{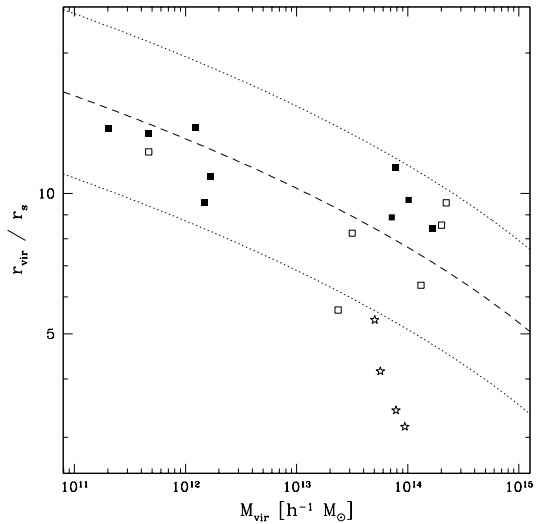}
  \caption{Relation between virial mass and NFW concentration
    parameter $c\equiv \Rv/r\sss$.
    Solid squares are used for relaxed systems, empty squares for
    minor mergers and stars for major mergers.
    Dashed line shows the model of \citet{Bullock01}.
    Dotted lines indicate the expected one-sigma scatter.
}
\label{figMc}
\end{figure}
%__________________________________

Another difference between merging and relaxed systems is that mergers
are typically more extended than relaxed haloes of the same mass.
We plot in Figure~\ref{figMc} the mass-concentration relation
for our sample, compared to the model of \citet{Bullock01},
\be
c(M,a)=K\frac{a}{a_c(M)}
\ee
where the collapse time $a_c$ is defined as the epoch at which the
typical collapsing mass, $M_*(a_c)$, equals a fixed fraction $F$ of the 
halo mass at epoch $a$.
According to \citet{Bullock01}, we set $K=3$ and $F=0.001$ in order to
account for the most massive haloes.
The scatter around the relation is fitted by $K=2$ and $K=4.7$.

Our sample of dark matter haloes is consistent with the expected
trend, although our cluster-size haloes seem to be slightly more
concentrated than the theoretical model.
On the other hand, major mergers display systematically
lower concentrations, since their collapse time is very close to the
present.
However, the value of $a_c$ is not well defined in a merging system.
In particular, during the first stages of the merger the profile
corresponds to an old object, perturbed by an approaching satellite.
It is only after virialisation when the profile relaxes to its final
form, and $a_c$ increases accordingly.

%___________________________________________________

\subsection{Angular momentum}

In addition to the radial mass distribution, the kinetic structure of
numerical haloes can offer interesting insights into the formation
of galaxies and galaxy clusters. In particular, we are interested in
the specific angular momentum of dark matter particles
in order to set the eccentricity in the analytical model.

We separate the velocity field into a random component
(i.e. velocity dispersion) and ordered rotation (i.e. average $j$).
The velocity dispersion of a collisionless CDM halo is
related to the total mass distribution by virtue of the Jeans
equation \citep[see e.g.][]{BT}.
For a spherically symmetric system with isotropic velocity
dispersion, neglecting infall,
\be
\frac{1}{\rho}\frac{\dd(\rho\sigma_r^2)}{\dd r}=-\frac{GM}{r}
\label{ecJeans}
\ee
where $\rho$ is the local density, $\sigma_r$ is the radial velocity
dispersion, and $M$ is the mass enclosed within radius $r$.
An approximate velocity dispersion profile could be derived by
substituting a given mass distribution (e.g. NFW)
and setting an arbitrary normalisation $\sigma_r(0)$.
The contribution of random motions to the angular momentum of the
CDM particles would be given by the tangential velocity dispersion.
In the isotropic case, this amounts to
$\left<j^2\right>=2r^2\sigma^2_r(r)$.
Note that random motions do not contribute to the total angular
momentum of the halo (i.e. $\left<j\right>=0$).

%__________________________________
\begin{figure}
  \centering \includegraphics[width=7cm]{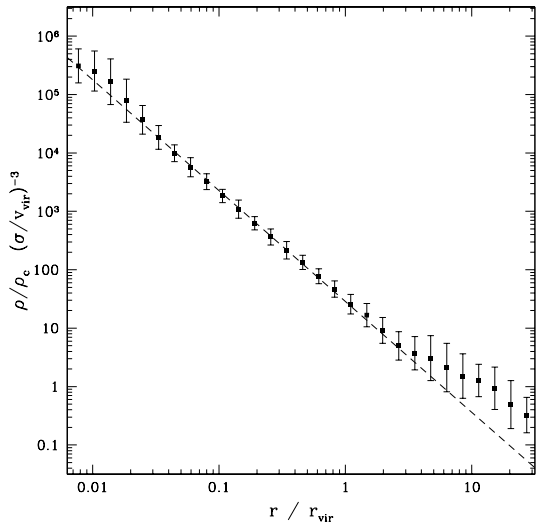}
  \caption{Phase-space density profiles of our haloes (dots).
    Error bars show the rms scatter of individual profiles around
    the average.
    The best-fit 'universal' profile (\ref{tnfit})
    is plotted as a dashed line.}
  \label{figTN01}
\end{figure}
%__________________________________

A more empirical approach to the velocity dispersion profile
has been followed by \citet{TN01}.
They realised that the coarse-grained phase-space density of a sample
of numerical galaxy-size haloes followed a power law
\be
\frac{\rho}{\sigma^3}\propto r^{\beta}
\label{ecTN01}
\ee
with $\beta=-1.875$ over more than two and a half decades in radius.
\citet{Rasia_03} obtained a similar result for cluster-size haloes,
although their best-fitting slope is $\beta=-1.95$.
The normalisation, though, has not been given in any of the two studies.

We have investigated the phase-space structure of the CDM haloes in
our sample, including both galaxies and galaxy clusters.
The average profile $\rho/\sigma^3$ is plotted in
Figure~\ref{figTN01}.
We find that all our haloes are well described by
\be
\frac{\rho}{\sigma^3} = 10^{1.46\pm0.04} \frac{\rho_c}{\Vv^3}
                        \left( \frac{r}{\Rv} \right)^{-1.90\pm0.05}
\label{tnfit}
\ee
where $\Vv^2\equiv GM_{\rm vir}/\Rv$.
This result is in fair agreement with the slopes
reported by \citet{TN01} and \citet{Rasia_03}.
The scatter around the average profile is remarkably low,
taking into account that our haloes span four orders of magnitude in
mass, and that we considered all systems (even major mergers)
in the analysis.
Indeed, we do not find any evidence that the slope of the relation
depends on mass or dynamical state.
Therefore, we claim that expression (\ref{tnfit}) can be regarded as a
'universal' phase-space density profile with only one free parameter,
which is the virial radius of the halo.

%__________________________________
\begin{figure}
  \centering \includegraphics[width=7cm]{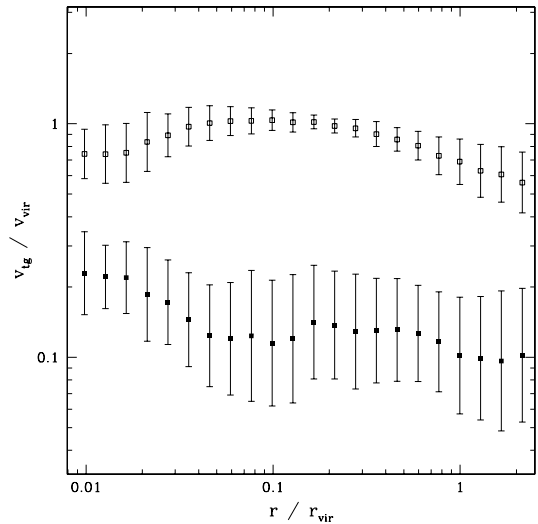}
  \caption{Contribution of bulk rotation (solid squares) and random
    motions (empty squares) to the tangential velocity
    of dark matter particles.
    Points represent the average over all haloes, and error bars
    indicate the rms scatter.
}
  \label{figVtg}
\end{figure}
%__________________________________

We now show in Figure~\ref{figVtg} that angular momentum
is indeed dominated by random motions.
We compare the contribution to the tangential velocity of CDM particles
from $<j>$ and $<j^2>$ over spherical shells.
The average bulk rotation velocity of our halos (solid symbols in
Figure~\ref{figVtg}) is about $0.1\Vv$,
although we find a significant increase near the centre.
Never the less, it is worth mentioning that the separation in bulk and
random velocity
is prone to numerical errors at the innermost regions.
The specific angular momentum grows roughly linearly with radius
(rigid body rotation), in agreement with previous numerical
work \citep{BarnesEfstathiou87,Bullock01,Bosch02,ChenJing02}.
In any case, the tangential velocity of individual dark matter
particles is always dominated by the velocity dispersion,
for all haloes, at any radius.

As was discussed in Section~\ref{secSCj}, the orbit eccentricity
is an important ingredient of our analytical model, since it
measures how deep a CDM particle is able to penetrate within the
gravitational potential.
The ideal case $e=0$ corresponds to a configuration in which all
particles would orbit at constant radius with no mixing between
concentric shells, while $e=1$ (often assumed in spherical infall
models) implies that every particle goes through the very centre of
mass of the halo.
For large values of the eccentricity, the central regions are
mainly composed of particles coming from the outer shells, that happen
to be near the pericentre of their orbits.

%__________________________________
\begin{figure}
  \centering \includegraphics[width=7cm]{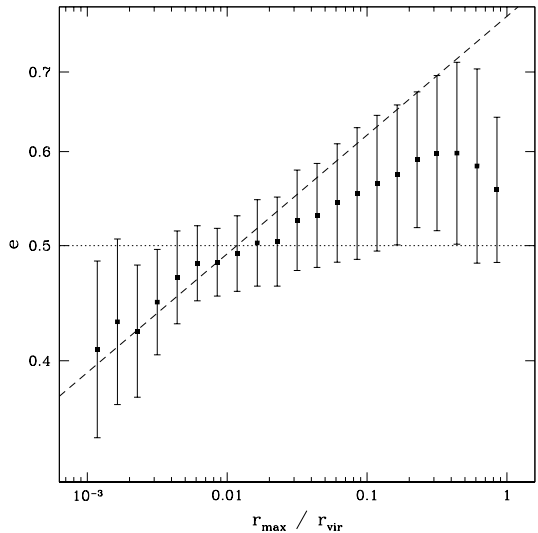}
  \caption{Orbit eccentricity of CDM particles as a function of apocentric
        radius. Dotted line marks $e=0.5$; dashed line,
        expression (\ref{ecEfit}).}
  \label{figE}
\end{figure}
%__________________________________

The eccentricity profile of our haloes is plotted in Figure~\ref{figE}.
For each object, we compute the spherically-symmetric
potential derived from its mass distribution.
The pericentric and apocentric radii of every CDM particle are estimated
from its position and velocity, and the eccentricity is computed as
\be
e=\frac{r_{\rm max}-r_{\rm min}}{r_{\rm max}+r_{\rm min}}.
\ee

%__________________________________
\begin{figure}
  \centering \includegraphics[width=8cm]{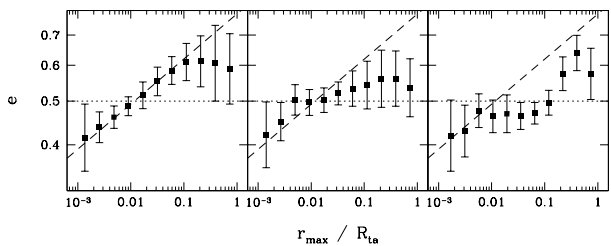}
  \caption{Same as Figure~\ref{figE}, separating relaxed haloes
    (left), minor (middle) and major (right) mergers.
}
  \label{figE2}
\end{figure}
%__________________________________

The general trend is that particle orbits are slightly more radial as
we move out to the current turn-around radius of the halo, $R_{\rm ta}$.
Moreover, we find a systematic dependence on the
dynamical state. As can be seen in Figure~\ref{figE2}, major mergers
are well described by constant eccentricity up to the virial radius
\footnote{For our dark matter haloes, $\Rv/R_{\rm ta}$ is typically of
  the order of $0.2-0.3$.},
while relaxed systems are more consistent with a power-law profile.
Minor mergers are somewhat in 
the middle. The average profile can be fitted by a power law, but the
slope is shallower than for relaxed systems.

A least-square fit to the relaxed population yields
\be
e(r_{\rm max})\simeq
0.8 \left( \frac{r_{\rm max}}{R_{\rm ta}} \right)^{0.1}
\label{ecEfit}
\ee
for $r_{\rm max}<0.1R_{\rm ta}$. We note that our approximation to
compute the eccentricity breaks down at larger radii, since the mass
distribution (and thus, the gravitational potential) are by no means
static beyond the virial radius.

%--------------------------------------------------------------------------
  \section{Model results}
  \label{secComp}
%--------------------------------------------------------------------------

We now attempt to reproduce the density profiles of our simulated
haloes with the model described in Section~\ref{secSC}.
In order to keep the number of free parameters to a minimum,
we assumed a constant eccentricity, $e\simeq0.5$,
which corresponds to the average over the whole radial range.
This value implies that particles are able to sink into
the dark matter potential up to one third of their maximum radius, or,
equivalently, that all mass within radius $r$ comes from shells whose
turn-around radius was $\le3r$.
The effect of a power law profile will be considered in
Section~\ref{secEpow}.

%___________________________________________________

\subsection{Constant eccentricity}

Once the eccentricity is set to $e=0.5$,
the model has two free
parameters, which describe the primordial density peak:
its height, $\nu$, and the smoothing scale, $R\ff$.
These parameters set the mass and formation time of the halo.
For a given mass, a higher peak on a smaller
scale corresponds to an earlier formation time and a steeper
density profile near the centre.

%__________________________________
\begin{figure}
  \centering \includegraphics[width=7cm]{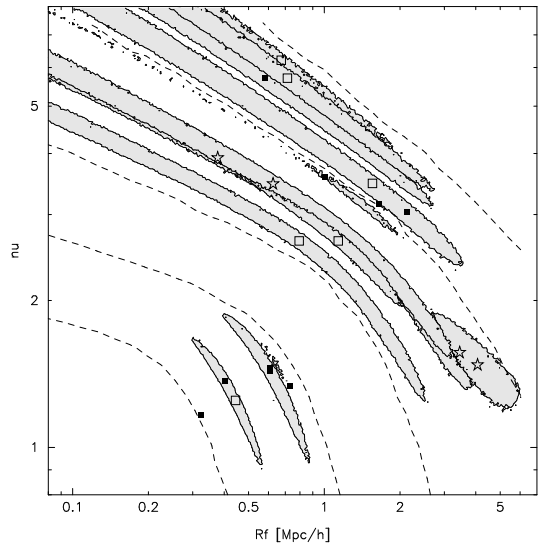}
  \caption{Best-fitting values of the peak height, $\nu$, and the
    smoothing scale, $R\ff$.
    Black squares represent relaxed haloes, open squares are used for
    minor mergers, and stars for major merging systems.
    Shaded areas indicate $\sqrt{\chi^2/(dof)}\le0.12$.
    Dashed lines are drawn at constant $\Rv$.
    }
  \label{figFitSC}
\end{figure}
%__________________________________

For each halo, we performed a $\chi^2$ minimisation of the mass profile.
Best-fitting values of $R\ff$ and $\nu$ are plotted in
Figure~\ref{figFitSC}, as well as the areas where
$\sqrt{\chi^2/(dof)}\le0.12$.
Most cluster-size haloes can be described as very high-$\nu$ peaks,
while galaxies seem to have collapsed around less extreme
fluctuations.
This means that clusters are expected to form
\emph{earlier} than galaxies, in the sense that the seeds of their
dark matter haloes are already in place at a higher redshift.

An exception to this rule are major mergers.
While some of these systems still have an early collapse time,
some others have collapsed approximately at the present epoch.
The first class corresponds to objects that have not relaxed yet,
and their core still corresponds to the old halo.
As relaxation completes and substructure is erased, the
best-fitting parameters move along the lines of constant
mass in the $\nu-R\ff$ diagram.
The smoothing scale rises sharply, and $\nu$ decreases accordingly.

As can be seen in Figure~\ref{figFitSC}, there is never the less a
certain degeneracy between $R\ff$ and $\nu$, which prevents a
reliable determination of the formation time. The exact value of the
best-fitting parameters may vary within the shaded area, depending on
the details of the fit. For instance, we tend to get higher peaks on
smaller scales as we give more weight to the inner parts of the profile.

%__________________________________
\begin{figure*}
  \centering \includegraphics[width=12cm]{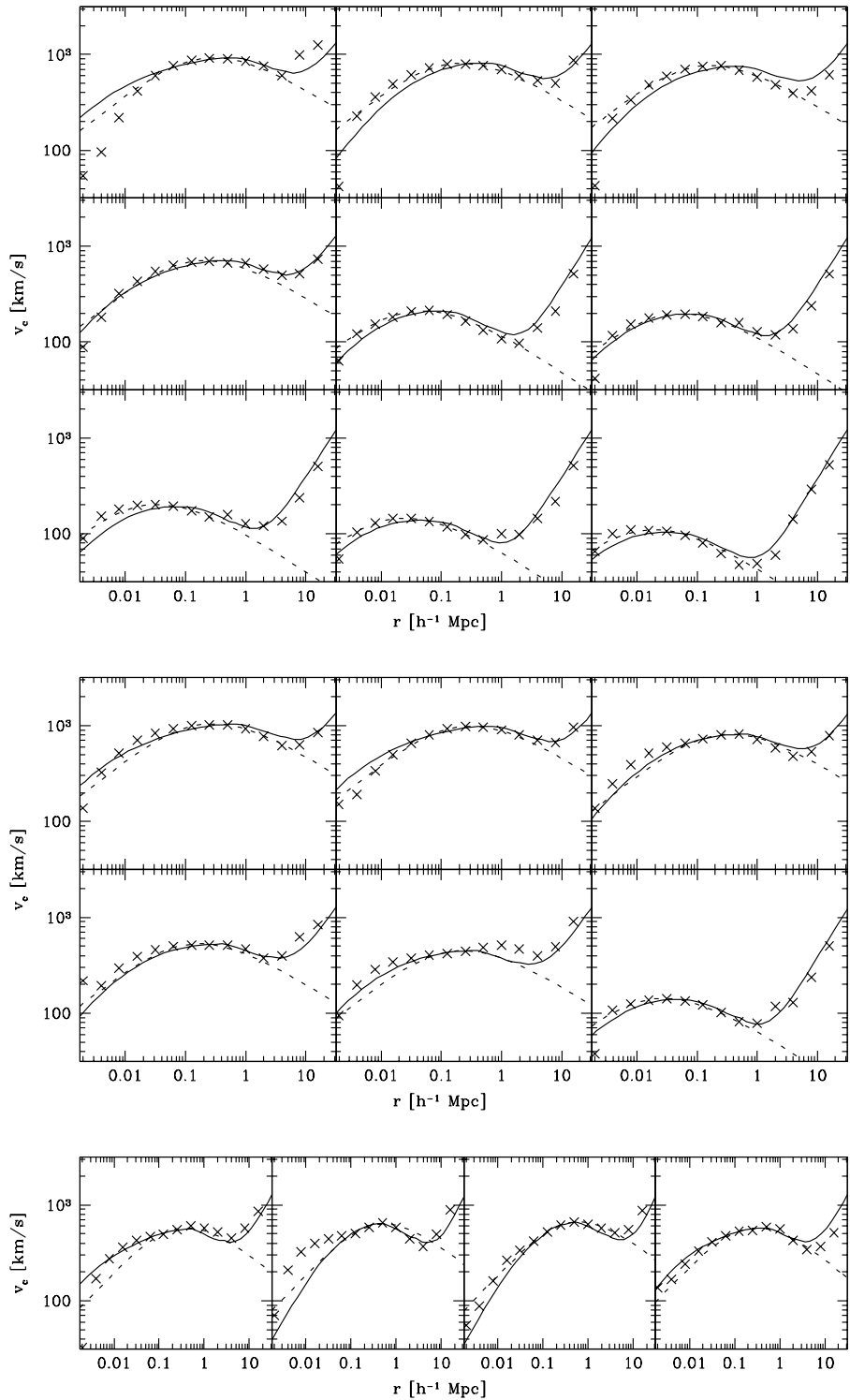}
  \caption{Circular velocity profiles, $v_c^2\equiv GM/r$.
    Crosses correspond to the simulation data, while dotted lines are the
    best-fitting NFW model and solid lines the best-fitting spherical 
    collapse model.
    Relaxed haloes, minor and major mergers are plotted on the top,
    middle and bottom panels, respectively.
    }
  \label{figVc}
\end{figure*}
%__________________________________

This can be understood when we compare the numerical data with the
results of our spherical collapse model.
We chose to fit the radial range $0.1<r/\Rv<1$ in order to test the
quality of the extrapolations, both towards the centre and to large radii.
Individual circular velocity profiles are shown in Figure~\ref{figVc},
together with the best fits provided by our model and NFW formula.

Although the mass distribution is fairly well described in general
terms, the central density is usually underestimated by both models.
When we fit the innermost parts, we are biased towards more
concentrated distributions. Indeed, if only data within
$0.1\Rv$ are considered, the best-fitting profiles are rather
unrealistic.

At large radii, the spherical collapse model reproduces the upturn in
the circular velocity, while NFW drops to zero. This is due to the
fact that the density in the NFW model vanishes at infinity, while it
tends asymptotically to the mean value in the spherical collapse
model. Adding a constant density background to the NFW
formula is enough to bring the profile in agreement with the
numerical data.

\begin{figure}
  \centering \includegraphics[width=8cm]{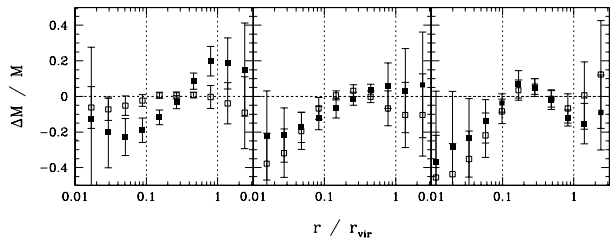}
  \caption{Accuracy of NFW (open squares) and spherical collapse
        (solid) models.
        Symbols correspond to the average over haloes classified as
        relaxed (left panel), minor (middle) or major mergers (right).
        Error bars indicate one-$\sigma$ scatter of individual
        profiles, and vertical dotted lines show the radial range used 
        for the fit.
}
  \label{figAccuSC}
\end{figure}

We assess the accuracy of both NFW and spherical collapse models in
Figure~\ref{figAccuSC}, where the quantity
\be
\frac{\Delta M}{M}\equiv\frac{M_{\rm model}-M_{\rm data}}{M_{\rm data}}
\ee
is plotted as a function of radius.

%__________________________________
%__________________________________

The most important difference between both models is that the accuracy 
of NFW fit improves significantly for relaxed systems, particularly
concerning the extrapolation towards $r\to 0$.
The fits based on our spherical collapse model are on average less
accurate, and the scatter between individual dark matter haloes is
a little bit larger.
Yet, the uncertainty is always lower than 20 per cent for relaxed
objects, and only slightly larger for merging systems, where the
prediction of the spherical collapse model is indeed very similar
to the NFW fit.

%___________________________________________________

\subsection{Variable eccentricity}
\label{secEpow}

Although constant eccentricity might provide a useful approximation for
merging systems, it is evident from Figure~\ref{figE2} that it is not
a good description of relaxed haloes, where the eccentricity profile
$e(r)$ increases significantly from the centre to the turn-around
radius. Moreover, the assumption of constant eccentricity leads to
systematic differences between the mass distribution predicted by our
model and the numerical data, as shown in Figure~\ref{figAccuSC}.

%__________________________________
\begin{figure}
  \centering \includegraphics[width=6cm]{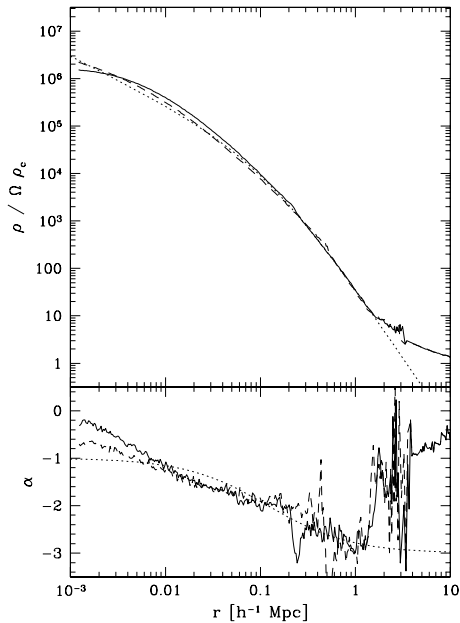}
  \caption{Comparison between a NFW density profile (dotted line),
    the spherical collapse model with constant eccentricity $e=0.5$
    (dashed line), and a power-law $e(r)$ given by
    (\ref{ecEfit}) (solid line).
    Top panel displays the density profile, and bottom panel its
    logarithmic slope.
}
  \label{figEpow}
\end{figure}
%__________________________________

Therefore, we would like to investigate the consequences of using a
more elaborate prescription for $e(r)$.
For constant eccentricity, high values of $e$ lead to steeper density
profiles in the central regions, whereas the outer parts of the halo
remain largely unaffected (see Figure~\ref{figSCj}).
We plot in Figure~\ref{figEpow} the density profile resulting from
our power-law fit (\ref{ecEfit}) for $\nu=3$ and $R\ff=1\ h^{-1}$
Mpc.
The mass distribution is very similar to that obtained with $e=0.5$,
but the shape is slightly different.
There is a small increase in the density between 10 and 100 $h^{-1}$
kpc, but the profile flattens near the centre due to the more circular 
orbits.

This flattening can be clearly seen on the bottom panel of
Figure~\ref{figEpow}, where we plot the logarithmic slope of the
density profile. It is interesting to note that the spherical collapse 
model predicts a finite density at $r=0$ when non-radial motions are
included, albeit the size of the 'core' is extremely small.
As was discussed in Section~\ref{secDM},
a similar trend is shown by the relaxed haloes in our sample.

%__________________________________
\begin{figure}
  \centering \includegraphics[width=8cm]{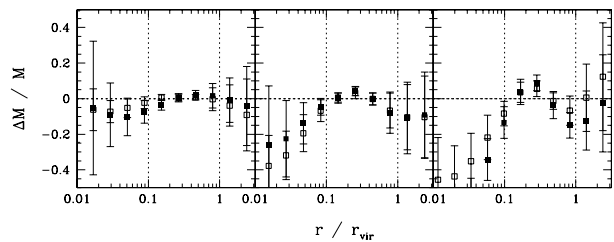}
  \caption{Same as Figure~\ref{figAccuSC}, when the eccentricity is
    set according to (\ref{ecEfit}).
}
  \label{figAccuEpow}
\end{figure}
%__________________________________

At our resolution limit, the net effect of using expression (\ref{ecEfit}) is
just to increase the density at small radii.
Then our model gives a somewhat poorer description of merging systems,
but the quality of the fit improves considerably for relaxed haloes.
As can be seen in Figure~\ref{figAccuEpow}, the accuracy of our model
is comparable to NFW formula when a realistic prescription is used for 
the eccentricity of particle orbits.

%--------------------------------------------------------------------------
  \section{Discussion and conclusions}
  \label{secConclus}
%--------------------------------------------------------------------------

The radial density profile of dark matter haloes has been
investigated within the framework of the spherical collapse theory.
We have shown that the model described in Section~\ref{secSC}
is able to reproduce the mass distribution of
realistic CDM haloes.
Although the final profile cannot be cast in a simple
analytical form, it provides not only
a mere phenomenological fit, but a
physically-motivated description of the density distribution
in terms of the primordial initial conditions.

However, it is not easy to understand how the assumption of spherical
symmetry could be able to describe the hierarchical assembly of
cosmological structures.
Instead of continuous infall of spherical shells,
the formation of CDM haloes observed in numerical
experiments takes place in a discrete and anisotropic way.
Most of the matter is accreted in clumps, along the preferred directions
set by the filamentary large scale structure.

None the less, the very complicated coalescence process looks very
regular in energy space \citep{Zaroubi96}.
Moreover, \citet{Moore99} have shown that the final density profile
is not very sensitive to the details of the merging history,
by comparing the mass distribution of a galaxy cluster halo
with a re-simulation in which the
power spectrum was truncated at $\sim4$ Mpc scales.

We suggest, as a plausible explanation for
the success of the spherical infall model, that merging is implicitly
taken into account through the smoothing scale $R\ff$.
Cosmological structures do not form around maxima of primordial
density field, but of the \emph{smoothed} density field \citep{HS85}.
Some memory of the initial conditions will be lost during major mergers,
particularly at the innermost regions of the resulting halo.
Contrary to the common view (see e.g. BBKS), we argue that
$R\ff$ has a precise physical interpretation;
below the mass scale defined by $R\ff$, all information about
the primordial substructure would have been erased by relaxation processes.

In the outer regions, matter is accreted in a more gentle way.
Minor mergers do not significantly alter the dynamical structure of the halo.
The mass of infalling clumps is much less than the average over a
spherical shell at large radii.
Thus, the density profile and the accretion rate are determined by the
amount of matter available to the halo,
which is ultimately set by the primordial initial conditions.

Never the less, there is still an additional degree of freedom,
related to the amount and distribution of angular momentum within
the dark matter halo.
Angular momentum sets the shape of the density profile at the inner regions.
For pure radial orbits, the core is dominated by particles from the outer
shells. As the angular momentum increases, these particles remain closer to
their maximum radius, resulting in a shallower density profile.

We found that angular momentum is dominated by the tangential component
of the velocity dispersion.
Random motions are well described by a 'universal' phase-space density
profile over several orders of magnitude in radius.
This profile is a power law with slope $\beta\simeq-1.9$,
in agreement with the results of \citet{TN01} for galactic-size haloes
and \citet{Rasia_03} for clusters.
We found that this profile is valid for all objects in our sample
(spanning a mass range from $10^{11}$ to $10^{15}$ M$_\odot$) as long 
as the virial radius is used as a scale parameter.

Although it would be desirable to establish a link between the initial
conditions and the specific angular momentum,
we resorted to a phenomenological approach.
According to the results of our numerical simulations, the orbit
eccentricity can be set to a constant value of $e=0.5$ as a first
order approximation. However, relaxed haloes are better described by a 
power-law profile in terms of their turn-around radius.
The details of the mass distribution depend only moderately on the
prescription assumed for the angular momentum.
Therefore, the spherical collapse model provides a valuable tool to
predict the structure of virialised dark matter haloes, given the
power spectrum of primordial fluctuations.
Indeed, we claim that the physical origin of the mass distribution
observed at the present day is related to the shape of the primordial
density peaks.
'Universal' profiles with two free parameters arise naturally from
Gaussian random peak statistics, since the primordial fluctuations are
fully specified by their height, $\nu$, and smoothing scale, $R\ff$.

%---------------------------------------------------------------

\section*{Acknowledgments}

The authors would like to thank Andreas Faltenbacher and Yehuda
Hoffman for useful discussions.
This work has been partially supported by the MCyT (Spain),
by the {\em Acciones Integradas Hispano-Alemanas} and
by {\em Deutscher Akademischer Austauschdienst} DAAD (Germany).
All simulations have been performed in the supercomputers of the
{\em Liebniz-Rechnenzentrum} LRZ (Germany).

%--------------------------------------------------------------------------
 \bibliographystyle{mn2e}
 \bibliography{../BibTeX/DATABASE,../BibTeX/PREPRINTS}

\end{document}